# Modulation of conductance and superconductivity by top-gating in LaAlO$_3$/SrTiO$_3$ 2-dimensional electron systems

P.D. Eerkes, W.G. van der Wiel and H. Hilgenkamp
*MESA+ Institute for Nanotechnology, University of Twente*
*P.O. Box 217, 7500 AE Enschede, The Netherlands*

We report the electrical top-gating of a 2-dimensional electron gas (2DEG) formed at the LaAlO$_3$/SrTiO$_3$ interface, using electron-beam evaporated Au gate electrodes. In these structures, epitaxial LaAlO$_3$ films grown by pulsed laser deposition induce the 2DEGs at the interface to the SrTiO$_3$ substrate and simultaneously act as the gate dielectric. The structured top-gates enable a local tuning and complete on/off switching of the interface (super)-conductivity, while maintaining the usual, intrinsic characteristics for these LaAlO$_3$/SrTiO$_3$ interfaces when no gate voltage is applied.

Conducting interfaces between insulating complex oxides exhibit appealing properties, both from a fundamental as well as from an applied perspective. A prominent example is the 2-dimensional electron gas (2DEG) at the interface between SrTiO$_3$ (STO) and LaAlO$_3$ (LAO).[1] Considerable low-temperature electron mobilities[1,2], superconductivity[3] and magnetic effects[4] have been reported, as well as a multiband/multiorbital character of the delocalized interface states, see e.g. [5], and the coexistence of multiple electronic phases[6-8]. These oxide interface systems show very good prospects for the realization of versatile and nano-scale[9] all-epitaxial devices, based on the electrostatic switching between different electronic/magnetic phases in transistor-like geometries.

Several groups, see e.g. [10-13], have reported successful back-gating experiments, where gate voltages are applied to the backside of the STO substrate at a distance of several hundreds of micrometers to the 2DEG. In this configuration, in spite of the large dielectric constant of STO, rather high gate voltages of up to several hundreds of volts are usually required to achieve sizable effects, due to the relatively high carrier concentrations of $10^{13}$-$10^{14}$ cm$^{-2}$ at the LAO/STO interfaces. Nevertheless, such back-gating experiments allowed for example for a complete on/off switching of superconductivity in a superconducting field effect transistor (SuFET) configuration[11], which has been a long pursued goal.[14] An important next development is to achieve top-gating close to the 2DEG, in which charge modulation can be performed on a local scale and also with smaller gate voltages, allowing more versatility in the realization of functional devices. Moreover, top-gating and back-gating can be employed simultaneously, allowing a control of the global and local conductance properties independently.

Förg *et al*.[15] realized all-oxide top-gated field-effect transistors using LAO/STO, and presented conductance modulation at temperatures from -100 °C to 100 °C. Their top-gates consisted of the high-$T_c$ superconductor YBa$_2$Cu$_3$O$_7$, which already at zero gate voltage reduced the 2DEG carrier density reduction to about 10% of its usual values. While this facilitates a subsequent full depletion of the interface with relatively low gate voltages, which is technologically very interesting, it also implies that the top-gating configuration modified the basic characteristics of the 2DEG. In this Letter, we describe top-gated structures using e-beam evaporated Au top-gate electrodes, which leaves the 2DEG properties unaffected. Local depletion of the 2DEG at room temperature and modulation of the superconducting properties at cryogenic temperatures are both achieved.

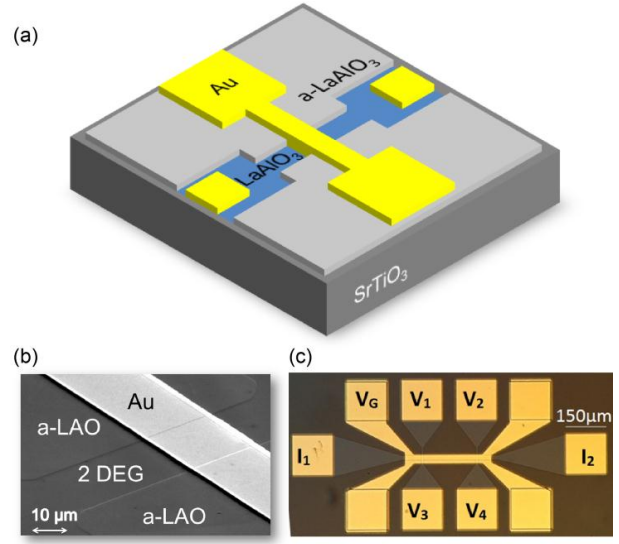

FIG. 1. (a) Schematic top view of the cross-bar structure. (b) Scanning electron micrograph of a cross-bar junction. (c) Optical micrograph of a top-gated Hall-bar structure.

The top gated LAO/STO devices were fabricated on SrTiO$_3$ (001)-oriented substrates, which were TiO$_2$-terminated according to the procedure described in Ref. [16]. To isolate the measurement structures, a lift-off patterned amorphous LaAlO$_3$ (a-LAO) layer was used.[17] The a-LAO was deposited by pulsed laser deposition (PLD) from a single-crystalline LAO target at room temperature in an oxygen pressure of $2\times10^{-3}$ mbar. Subsequently, crystalline LAO was PLD-deposited in an oxygen pressure of $2\times10^{-3}$ mbar and at 850 °C. The growth of this layer was monitored by *in situ* reflection high-energy electron diffraction (RHEED). After deposition, the sample was cooled down to 600 °C at deposition pressure. Subsequently, a post anneal of 1 hour at 600 °C at an oxygen pressure of 600 mbar was applied, followed by a cool down to room temperature at this pressure. This procedure renders the a-LAO/STO areas fully insulating, as intended. If no post anneal is applied, the room-temperature sheet resistivity of these areas is around 1 M$\Omega$/$\square$. With the post anneal, it is above the measurement limit of 50 G$\Omega$/$\square$. Ohmic contacts to the 2DEG were deposited by high-power sputtering of Ti/Au. As a final step, e-beam evaporation of 100 nm Au (without Ti) was used for the deposition of the top-gate, patterned by lift off. This deposition technique ensures a relatively soft landing of the Au atoms at the LAO surface (compared to sputtering), resulting in low leakage currents as shown below. We note that if instead sputtered Au gate electrodes were used, the gates were always rather leaky.

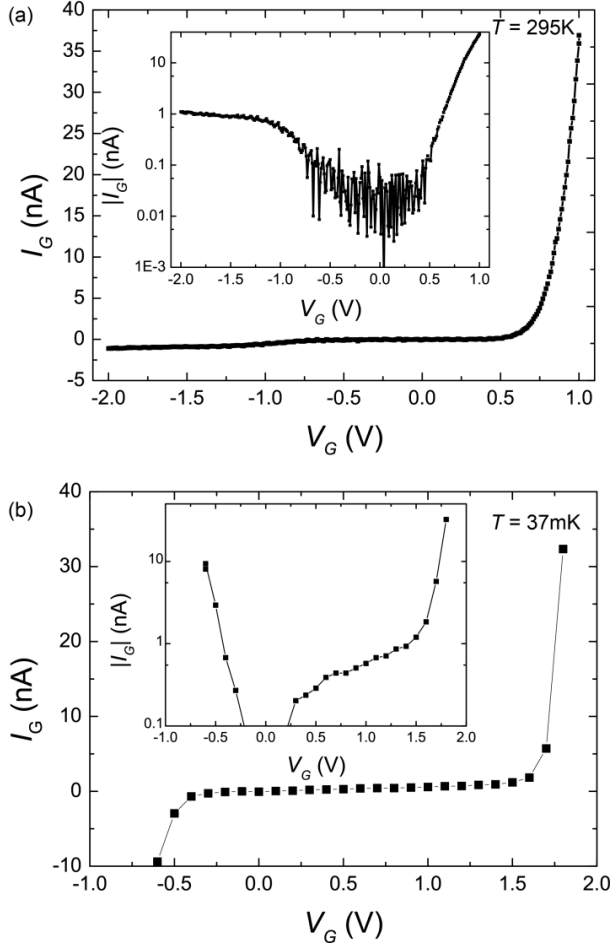

FIG. 2. Gate current $I_G$ as a function of gate voltage $V_G$ measured at $I_{SD} = 0$. The insets show the absolute value of the gate current on a logarithmic scale. (a) Room temperature measurement for an 11 u.c. LAO cross-bar junction with a gate-area of 500 μm². (b) Gate current for a 12 u.c. LAO/STO structure at 37 mK. The effective gate-area in this device is 5040 μm².

The 2DEGs were patterned in two different structures. Figure 1a shows a schematic of a cross-bar junction. First, a strip is structured in the LAO/STO 2DEG system. Crossing this, a Au bridge was fabricated, acting as the top-gate (Fig. 1b). Figure 1c shows a 15 μm wide Hall-bar structure. The source-drain current $I_{SD}$ is applied between $I_1$ and $I_2$ (ground). The 2DEG resistance is measured between voltage probes $V_1$ and $V_2$ or between $V_3$ and $V_4$ (200 μm separation) and the gate-voltage $V_G$ was always applied with respect to ground. The total 2DEG area under the gate-electrode in this device was 5040 μm². Various samples have been fabricated this way, showing comparable characteristics. In this Letter we will focus on two devices grown in different deposition runs.

The first is a cross-bar structure with a 2DEG width of 25 μm and a width of the Au top-gate of 20 μm. This sample, which we measured in detail at room temperature, comprised of 11 unit cells (u.c.) of LAO, which corresponds to 4.1 nm. The second was a Hall-bar structure, with dimensions as in Fig. 1c and containing 12 u.c. (4.5 nm) of LAO. This was investigated more extensively at low temperatures.

For both devices, first the gate current $I_G$ was investigated. Figure 2a shows $I_G$ as function of $V_G$ for the cross-bar junction at room temperature. For negative top-gate voltages up to −2 V, $I_G$ did not significantly exceed ~1 nA. For positive gate voltages, $I_G$ remained low up to about $V_G = 0.5$ V, increasing above this voltage to $I_G = \sim 40$ nA at $V_G = 1$ V. Figure 2b shows $I_G(V_G)$ of the Hall-bar at 37 mK. The gate current for this rather large structure remained below 10 nA for gate voltages between -0.6 and +1.7 V. It is noted that while one sees an increase in the gate current for negative gate voltages exceeding about -0.4 V, at room temperature gate-voltages up to -200 V could be applied without a significant gate current. For $V_G < -0.6$ V, the 2DEG became completely insulating. Conductance could then only be recovered when warming the devices up to about 160 K. This is a similar temperature as reported before needed to restore the initial cool down conductivity.[18,19]

Figure 3a shows $I_{SD}$ as a function of $V_G$ for several $V_{SD}$-values measured at room temperature for the cross-bar junction on the 11 u.c. LAO sample. A gate voltage of -1 V is sufficient for this sample to locally deplete the conducting channel below the gold top-gate, pinching off the source-drain current to residual values < 1nA. Figure 3b shows $I_{SD}(V_{SD})$ curves for several gate voltages of the same junction.

To verify that the presence of the Au top-gates does not influence the total amount of charge carriers in the idle state, $V_G = 0$ V, Hall measurements were performed at 2 K in the Hall-bar structure. The Hall voltage was measured between contacts $V_1$ and $V_3$. From the Hall signal, we calculate the carrier density $n_s$, in a one-band approximation. For $V_G = 0$ V, we find $n_s = 1.9 \times 10^{13}$ cm$^{-2}$, which equals the typical values recorded for similar interfaces without the metallic top-gates.[20] From the slope of $n_s$ as a function of $V_G$ (Fig. 4), and assuming a simple parallel-plate-capacitor model, the dielectric constant of the thin LAO layer is estimated to be around 7 at this measurement temperature of 2 K. This is lower than the reported bulk value of 25, which may partly be related to an underestimate of the capacitor thickness in the parallel plate model, due to the distance over which the 2DEG extends into the SrTiO$_3$ substrate.

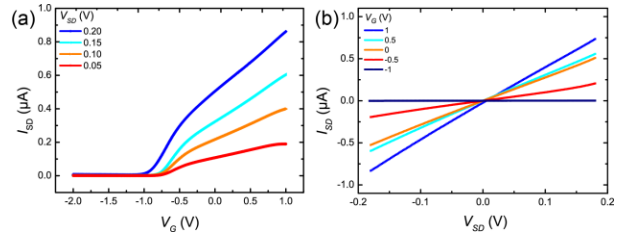

FIG. 3. (a) Drain-source current $I_{DS}$ as a function of top-gate voltage for source-drain voltages of 50, 100, 150 and 200 mV for an 11 u.c. LAO cross-bar structure with a gate area of 500μm². (b) $I_{SD}(V_{SD})$ curves of the same structure for top-gate voltages of -1, -0.5, 0, 0.5 and 1 V. Both measurements are at room temperature.

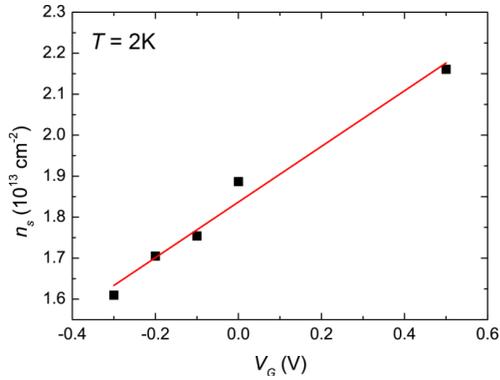

FIG. 4. Carrier density as a function of gate voltage for the sample of Fig. 4, measured at 2 K. The carrier density at 2 K of $1.9 \times 10^{13}$ cm$^{-2}$ for $V_G = 0$V is comparable to LAO/STO 2DEGs grown under similar conditions but without a top-gate.

To investigate the superconducting properties, devices with Hall-bar structures were cooled down in a dilution refrigerator. No special cool down procedure was performed other than that all contacts were grounded until the measurement temperature of 37 mK was reached. In Fig. 5 we show $I_{SD}(V_{SD})$ curves of a sample with a 12 u.c. LAO layer thickness, for different $V_G$- values ranging from -0.6 V to +2.0 V in steps of 0.2 V. The sample shows superconductivity below approx. 240 mK if no voltage is applied to the gate. A positive $V_G$ raises the critical temperature to about 275 mK at $V_G = 1.4$ V, simultaneously increasing the critical current for superconductivity at 37 mK (Fig. 5). At a gate voltage of +1.5 V, the gate current rapidly increases and the associated energy dissipation suppresses the superconductivity in the LAO/STO interface. Upon applying a negative gate voltage, the critical current is gradually decreased. To our knowledge, this result presents the first demonstration of a low-gate voltage SuFET device in a top-gated all-oxide 2DEG system.

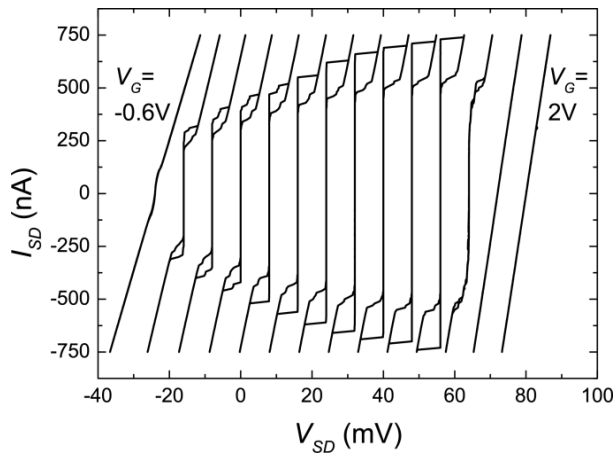

FIG. 5. $I_{SD}(V_{SD})$ - curves of a superconducting LAO/STO Hall-bar with 12 u.c. of LAO, measured at 37 mK for top-gate voltages from -0.6V to +2.0V in steps of 0.2V. The curves are offset for clarity.

In summary, we fabricated structured LAO/STO devices with metallic top-gates. Low leakage currents are obtained for the entire temperature range between 37 mK and 300 K, and in the idle state ($V_G = 0$ V) the characteristics of the 2DEGs resemble those of devices without gates. Both at room temperature as well as at cryogenic temperatures the (super)conductance can be completely suppressed, which gives tantalizing prospects for the creation of quantum/nano structures analogous to the gated structures investigated in semiconductor 2DEG systems. The electronic correlations and associated rich phase diagrams in the complex oxides are anticipated to lead to novel characteristics for such devices, with intriguing potential for applications.

**Acknowledgments**
This research was supported by the Dutch Foundation for Research on Matter through the InterPhase program. W.G.v.d.W. acknowledges financial support from the European Research Council, ERC StG. no. 240433. The authors thank A. Brinkman, M. Huijben, G. Koster, M. Kruize, X. Renshaw Wang, G. Rijnders, and S. Wenderich for valuable discussions.

[1] A. Ohtomo and H. Y. Hwang, Nature **427** (6973), 423 (2004).
[2] M. Huijben, G. Koster, M.K. Kruize, S. Wenderich, J. Verbeeck, S. Bals, E. Slooten, B. Shi, H.J.A. Molegraaf, J. E. Kleibeuker, S. van Aert, J.B. Goedkoop, A. Brinkman, D.H.A. Blank, M.S. Golden, G. van Tendeloo, H. Hilgenkamp and G. Rijnders, Advanced Functional Materials, DOI: 10.1002/adfm.201203355 (2013).
[3] N. Reyren, S. Thiel, A. D. Caviglia, L.F. Kourkoutis, G. Hammerl, C. Richter, C.W. Schneider, T. Kopp, A.S. Ruetschi, D. Jaccard, M. Gabay, D.A. Muller, J.M. Triscone and J. Mannhart, Science **317** (5842), 1196 (2007).
[4] A. Brinkman, M. Huijben, M. van Zalk, J. Huijben, U. Zeitler, J.C. Maan, W.G. van der Wiel, G. Rijnders, D.H.A Blank and H. Hilgenkamp, Nature Materials **6** (7), 493 (2007).
[5] S.W.A. McCollam, M.K. Kruize, V.K. Guduru, H.J.A. Molegraaf, M. Huijben, G. Koster, D.H.A. Blank, G. Rijnders, A. Brinkman, H. Hilgenkamp, U. Zeitler and J.C. Maan, Arxiv **1207.7003v1** (2012).
[6] Ariando, X. Wang, G. Baskaran, Z.Q. Liu, J. Huijben, J.B. Yi, A. Annadi, A.R. Barman, A. Rusydi, S. Dhar, Y.P. Feng, J. Ding, H. Hilgenkamp and T. Venkatesan, Nature Communications **2**, 188 (2011).
[7] D.A. Dikin, M. Mehta, C.W. Bark, C.M. Folkman, C.B. Eom and V. Chandrasekhar, Physical Review Letters **107**, 056802 (2011).
[8] J. A. Bert, B. Kalisky, C. Bell, M. Kim, Y. Hikita, H.Y. Hwang and K. A. Moler, Nature Physics **7** (10), 767 (2011).
[9] C. Cen, S. Thiel, J. Mannhart and J. Levy, Science **323**, 1026 (2009).
[10] S. Thiel, G. Hammerl, A. Schmehl, C.W. Schneider and J. Mannhart, Science **313**, 1942 (2006).
[11] A.D. Caviglia, S. Gariglio, N. Reyren, D. Jaccard, T. Schneider, M. Gabay, S. Thiel, G. Hammerl, J. Mannhart and J. M. Triscone, Nature **456**, 624 (2008).
[12] C. Bell, S. Harashima, Y. Kozuka, M. Kim, B. G. Kim, Y. Hikita and H.Y. Hwang, Physical Review Letters **103**, 226802 (2009).
[13] M. Ben Shalom, M. Sachs, D. Rakhmilevitch, A. Palevski and Y. Dagan, Physical Review Letters **104**, 126802 (2010).
[14] J. Mannhart, Supercond. Sci. Technol. **9**, 49 (1996).
[15] B. Förg, C. Richter and J. Mannhart, Applied Physics Letters **100**, 053506 (2012).
[16] G. Koster, B.L. Kropman, G.J.H.M. Rijnders, D.H.A. Blank, and H. Rogalla, Applied Physics Letters **73**, 2920 (1998).
[17] C.W. Schneider, S. Thiel, G. Hammerl, C. Richter and J. Mannhart, Applied Physics Letters **89**, 122101 (2006).
[18] S. Seri, M. Schultz and L. Klein, Physical Review B **87**, 125110 (2013).
[19] S.H.J. Biscaras, C. Feuillet-Palma, A. Rastogi, R.C. Budhani, N. Reyren, E. Lesne, D. LeBoeuf, C. Proust, J. Lesueur, N. Bergeal, Arxiv **1206.1198v1** (2012).
[20] M. Huijben, A. Brinkman, G. Koster, G. Rijnders, H. Hilgenkamp and D.H.A. Blank, Advanced Materials **21**, 1665 (2009).